\documentclass{wileyarticle}

\graphicspath{{images/}}
\usepackage{aas_macros}
\articletype{}%
\artid{}%
\jid{}%
\jvol{}
\jyear{}%
\jissue{}%
\jmonth{}%
\doi{}%
\copyeditor{}%
\startpage{1}%
\received{}
\revised{}
\accepted{}

\begin{document}

\title{Magneto-elastic oscillations modulating the emission of magnetars}

\author[1]{Michael Gabler}

\author[2]{Pablo Cerd\'a-Dur\'an}

\author[1]{Alexei Mate}

\author[3]{Nikolaos Stergioulas}

\author[2,4]{Jos\'e A.~Font}

\author[1]{Ewald M\"uller}

\authormark{Gabler \textsc{et al}}

\address[1]{Max-Planck-Institut f\"ur Astrophysik,
  Karl-Schwarzschild-Str.~1, 85741 Garching, Germany}

\address[2]{Departamento de Astronom\'{\i}a y Astrof\'{\i}sica,
  Universidad de Valencia, 46100 Burjassot (Valencia), Spain}

\address[3]{Department of Physics, Aristotle University of Thessaloniki,
  Thessaloniki 54124, Greece}
\address[4]{Observatori Astron\`omic, Universitat de Val\`encia, C/ 
Catedr\'atico Jos\'e Beltr\'an 2, 46980, Paterna (Val\`encia), Spain}

\corres{Michael Gabler, Karl-Schwarzschild-Str.~1, 85741 
Garching, \email{miga@mpa-garching.mpg.de}}

\begin{abstract}%

Magneto-elastic oscillations of neutron stars are believed to explain observed 
quasi-periodic oscillations (QPOs) in the decaying tail of the giant flares of 
highly magnetized neutron stars (magnetars). Strong efforts of the 
theoretical modelling from different groups have increased our understanding of 
this phenomenon significantly.
Here, we discuss some constraints on the matter in neutron stars that 
arise if the interpretation of the observations in terms of superfluid, 
magneto-elastic oscillations is correct.
To explain the observed modulation of the light curve of the giant 
flare, we describe a model that allows the QPOs to couple to the 
stellar exterior through the magnetic field. In this magnetosphere, the shaking 
magnetic field induces currents that provide scattering targets for 
resonant cyclotron scattering of photons, which is calculated with 
a Monte-Carlo approach and coupled to a code that calculates the momentum 
distribution of the charge carriers as a one-dimensional accelerator problem. 
We show first results of a simplified, but self-consistent momentum 
distribution, i.e. a waterbag distribution, and of the corresponding spectra.

\end{abstract}


\maketitle

\section{Introduction}\label{sec_intro}
The quasi-periodic oscillations (QPOs) observed in the giant flares of the
magnetars SGR 1806-20 and SGR 1900+14, 
respectively 
\citep[see e.g.][and references 
therein]{Israel2005,Strohmayer2005,Watts2006,Strohmayer2006,Huppenkothen2014c} 
,
are commonly interpreted as torsional neutron star ocillations. Different 
groups have made strong efforts to understand these magneto-elastic 
oscillations  \citep[see e.g.][and references 
therein]{Duncan1998, Piro2005, Sotani2007, 
Samuelsson2007, Steiner2009,Cerda2009, Sotani2008, 
Colaiuda2009,Levin2006,Levin2007,Glampedakis2006b, Gabler2011letter, 
Gabler2012, Colaiuda2011, vanHoven2011, vanHoven2012,Passamonti2013, 
Gabler2013letter, Passamonti2014, Gabler2016}.
These theoretical models have reached a very sophisticated level and are able 
to explain most of the frequencies, at least in principle. However, there are 
degeneracies in the parameter space between equation of state (EoS), mass or 
compactness of the particular neutron star model, magnetic field strength and 
configuration, and assumptions about superfluidity, that make it hard to 
directly associate the observed frequencies to a particular model and, thus, to 
constrain the coresponding model properties. 

Furthermore, there exists no satisfactory model to explain the modulation of 
the 
emission process. \cite{Timokhin2008} suggested that resonant 
cyclotron scattering (RCS) of photons emitted from the surface by charged 
particles in the magnetosphere have the potential to modulate the 
light curve of the giant flare. This idea is based on models by 
\cite{Thompson2002, 
Lyutikov2006, Nobili2008, Fernandez2007, Beloborodov2007} who explain the 
non-flaring state with the RCS model. Active magnetars, 
i.e. magnetars  with hard X-ray components, posses a twisted magnetosphere, 
which requires currents to maintain its twist. 
These currents are formed by electrons and positrons and they can 
scatter the 
photons emitted from the surface of the neutron star. Due to this interaction 
the photons are upscattered in energy by $\gamma^2$, where $\gamma$ is the 
Lorentz factor of the 
scattering 
particles \citep{Beloborodov2013long}, and the spectrum changes. The 
calculation 
of the momenta of the charge carriers is thus an essential ingredient to 
determine the photon properties after scattering. First studies assumed mildly 
relativstic flows \citep{Nobili2008, Fernandez2007}, but it was shown 
that 
this is an unrealistic assumption, because the particles are highly 
relativistic and interact strongly with the photons creating a self-regulated 
flow 
\citep{Beloborodov2007, Beloborodov2009,
Beloborodov2013long, Beloborodov2013, Hascoet2014, Chen2017}. 

As a first step to self-consistently describe how the torsional oscillations of 
the neutron star can modulate the emission, we showed how these 
oscillations 
shake the magnetosphere by steadily twisting and untwisting the exterior 
magnetic field \citep{Gabler2014}. With the simplifying assumption that the 
required currents are conducted by mildly relativistic charge carriers, we also 
showed that surface amplitudes of the oscillations of less than 
$1\,$km are sufficient to significantly modulate the light curve in 
the energy band 
where the QPOs were observed \citep{Gabler2014Proceedings}.

Here, we summarize the theoretical model of magneto-elastic oscillations of 
magnetars in Section\,\ref{sec_QPO} and discuss how to constrain
 properties of high density matter. In Section\,\ref{sec_emission} we 
present our advance in modeling the RCS in magnetar magnetospheres, and in 
Section\,\ref{sec_conclusion} we summarize our model explaining QPOs in the 
giant flares of magnetars.

\section{Magneto-elastic oscillations}\label{sec_QPO}
In previous work, we studied how the oscillations are influenced by 
different 
magnetic field configurations and magnetic field strengths \citep{Gabler2013}, 
different 
equations of state, different neutron star masses\citep{Gabler2012}, 
and superfluid parameters \citep{Gabler2013letter,Gabler2016}.
According to these studies, there are two general conditions that can be 
used to obtain constraints on the properties of high density matter:
(i) The oscillations have to reach the surface and (ii) there has to 
be a high frequency oscillation, i.e. a resonance between a crustal shear mode 
and a high Alfv\'en overtone of the core 
\citep{Gabler2013letter, Passamonti2014, Gabler2016}.
As we will show in an accompanying paper \cite{Gabler2017} 
these 
two conditions can be cast into the following equations (without relativistic 
corrections):
\begin{equation}
\bar B_{14} = 17.23 
\sqrt{\varepsilon_* X_c} \sqrt{\frac{\mu_{cc}}{\mu_{cc,{\rm ref}}}} . 
\label{eq:bbreak}
\end{equation}
and
\begin{equation}
 \mu_{cc}\lesssim \frac{4f_\mathrm{obs}^2 \Delta r^2\rho_{cc}}{n^2(1+a_{2t_n} 
\bar B_{14}^2)}\,,\label{eq:n1}
\end{equation}
respectively. Here, we introduced the mass fraction of charged particles $X_c$, 
the entrainment factor $\varepsilon_*$,  the shear modulus $\mu_{cc}$, the
 density $\rho_{cc}$ at the core-crust interface, the crust size $\Delta 
r$, $\bar B_{14}=\bar B / 10^{14}\,$G, $\mu_\mathrm{cc,ref}=2.09 \times 
10^{30}\,$erg/cm$^3$ and a fitting factor $a_{2t_n}$ derived 
in \cite{Gabler2017}. The compactness of our fiducial model is $M/R=0.1687$.

\begin{figure}
\SPIFIG{\includegraphics[width=.48\textwidth]{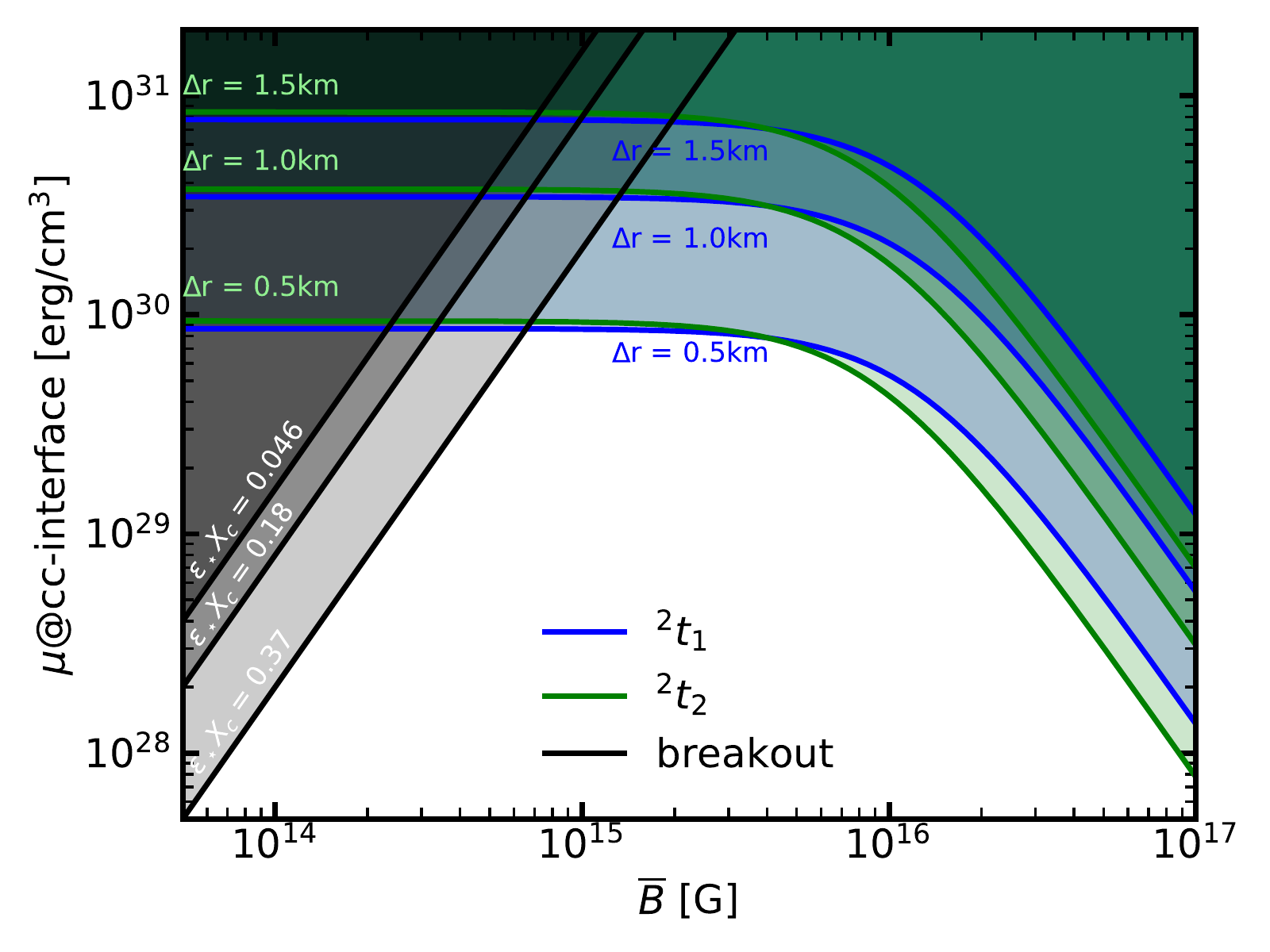}}
 \caption{Constraints in the $\mu_\mathrm{cc}- \bar B$ plane. Black lines 
indicate the magnetic field for which the oscillations reach the surface, and 
blue (green) lines give the limit at which the frequency of the $^2t_1$ 
($^2t_2$) crustal overtone is too high. All shaded areas are 
disfavored in our 
model.}\label{fig_mu_B}
\end{figure}

In Figure\,\ref{fig_mu_B} we plot the constraints in the $\mu_{cc}-\bar B$ 
plane. The black lines indicate the minimal $\bar B$ for a given $\mu_{cc}$ at 
which the QPOs can reach the surface given by Eq.\,(\ref{eq:bbreak}). Only 
magnetic fields this line are allowed in our model. Shaded 
regions are disfavored. $\bar B$ depends sensitively on 
the assumed value of the superfluid parameters, because the reflection at the 
core-crust interface depends on the jump in propagation speeds of a perturbation 
crossing from one side of the interface to the other \citep{Gabler2012}. This 
jump decreases with 
an increasing fraction of superfluid neutrons in the core
\citep{Gabler2016}. Therefore, the threshold for the breakout for a given 
$\mu_{cc}$ decreases with decreasing $\varepsilon_* X_c$. The blue (green) 
curves originate from the presence of a high frequency resonance with the 
$^2t_1$ ($^2t_2$) crustal 
overtone that should have a frequency lower than $625\,$Hz ($1840\,$Hz). 
Depending on the size of the crust, this condition provides a constraint on the 
shear modulus $\mu_{cc}$. We show exemplarily curves for $\Delta r=\{0.5, 
1.0, 
1.5\}\,$km. For a larger crust, the shear modulus can be larger. 
From Fig.\,\ref{fig_mu_B} and by assuming a shear modulus of 
$\mu_{cc}=8\times10^{29}\,$erg/cm$^3$, we can see, that the magnetic field has 
to be 
in the range of $\bar B\gtrsim2\times10^{14}\,$G (breakout at $\varepsilon_* 
X_c=0.046$)  to $\bar B \lesssim 
5\times10^{15}\,$G 
($^2t_1$) and that the crust thickness should be larger than $\Delta 
r\gtrsim0.5\,$km.

\begin{figure}
 \SPIFIG{\includegraphics[width=.48\textwidth]{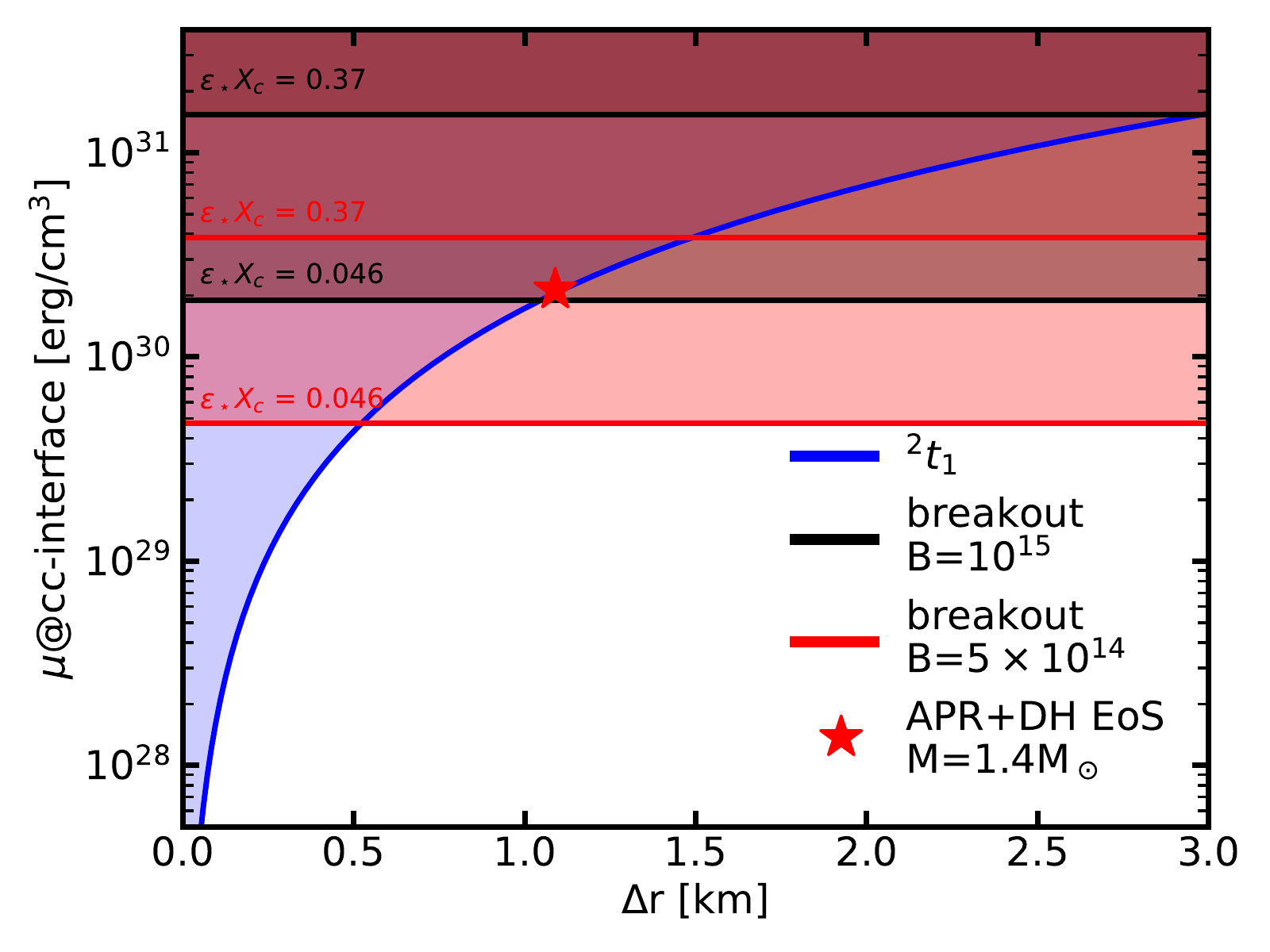}}
 {\caption{Constraints in the $\mu_\mathrm{cc}-\Delta r$ plane. The blue line 
indicates the threshold for the $^2t_1$ oscillation having a frequency below 
$625\,$Hz} and the black (red) lines give the limit for the breakout of the 
oscillations to the surface at $\bar B=10^{15}\,$G ($\bar B=5\times10^{14}\,$G)}
\label{fig_mu_DR}
\end{figure}

In Fig.\,\ref{fig_mu_DR}, we use Eq.\,(\ref{eq:bbreak}) (red and black lines) 
and 
(\ref{eq:n1}) with $\bar B_{14}=0$ as lower limit (blue line)  to constrain the 
parameters in the $\mu_{cc} - \Delta 
r$ - plane. For the former we show two exemplary values of the entrainment 
$\varepsilon_* X_c=0.046$ and $0.37$ and two magnetic field strengths $\bar 
B=5\times10^{14}$ and $10^{15}\,$G. As before, shaded regions are disfavored 
by our 
model. For a shear modulus of $\mu_{cc}=10^{30}$erg/cm$^3$, the crust should 
be larger than $\Delta r=0.7\,$km. The breakout only constrains the possible 
shear moduli, but does not limit $\Delta r$. 

\section{Modulating the magnetar emission}\label{sec_emission}
As explained before, the theoretical modelling of the oscillations inside the 
neutron star has received great attention. However, there is yet no 
self-consistent description of a mechanism that can cause the observed 
modulation of the light curve of the giant flare. In a first paper, we studied 
how the oscillations can modify and shake the exterior magnetic field 
\citep{Gabler2014}, and we found that only oscillations symmetric in $\delta 
B_\varphi$ can be excited outside the neutron star.

Following the idea of \cite{Timokhin2008}, we investigate the possibility of 
photons being upscattered by resonant cyclotron scattering (RCS). The basic 
model requires currents in the magnetosphere, which, in our case, are generated 
by the steady twisting and untwisting of the magnetic field due to the internal 
oscillations. For a consistent calculation of the resulting change in 
the spectrum of the photons, one needs to know the momentum distribution of the 
charge carriers of the currents. We describe the e$^\pm$ pair plasma with a 
waterbag distribution characterized by the maximum momentum of positrons $p_+$. 
This plasma strongly interacts with the photons in a way described by  
\cite{Beloborodov2007} and
\cite{Beloborodov2009,Beloborodov2013long} and it
self-regulates 
towards a quasi steady state: The particles accelerate along the magnetic 
field lines due to the twisted magnetic field which causes a potential 
difference between the footpoints of the field line. Then the e$^\pm$ decelerate 
due to the interaction with photons, which get reflected in a region close to 
the equator. The photons encounter an almost opaque region there, because the 
charge carriers are slowed down significantly due to their interaction with 
photons. However, the e$^\pm$ still have to conduct 
the current required by the twisted field. Thus, the 
density increases significantly creating a huge optical depth for photons.
We iteratively calculate this interaction between photons, which slow down the 
charge carriers, and the latter, which scatter the former.
In a first step we calculate the drag force a radially streaming photon field 
would exert on the $e^\pm$ flowing along the field lines. The resulting 
currents with the calculated momentum distribution, are used in a second step 
to calculate the scattering of photons with a Monte Carlo method. After a 
few iterations repeating this procedure, both the spectrum and momentum 
distribution of charge carriers converge.

\begin{figure}
 \SPIFIG{\includegraphics[width=.48\textwidth]{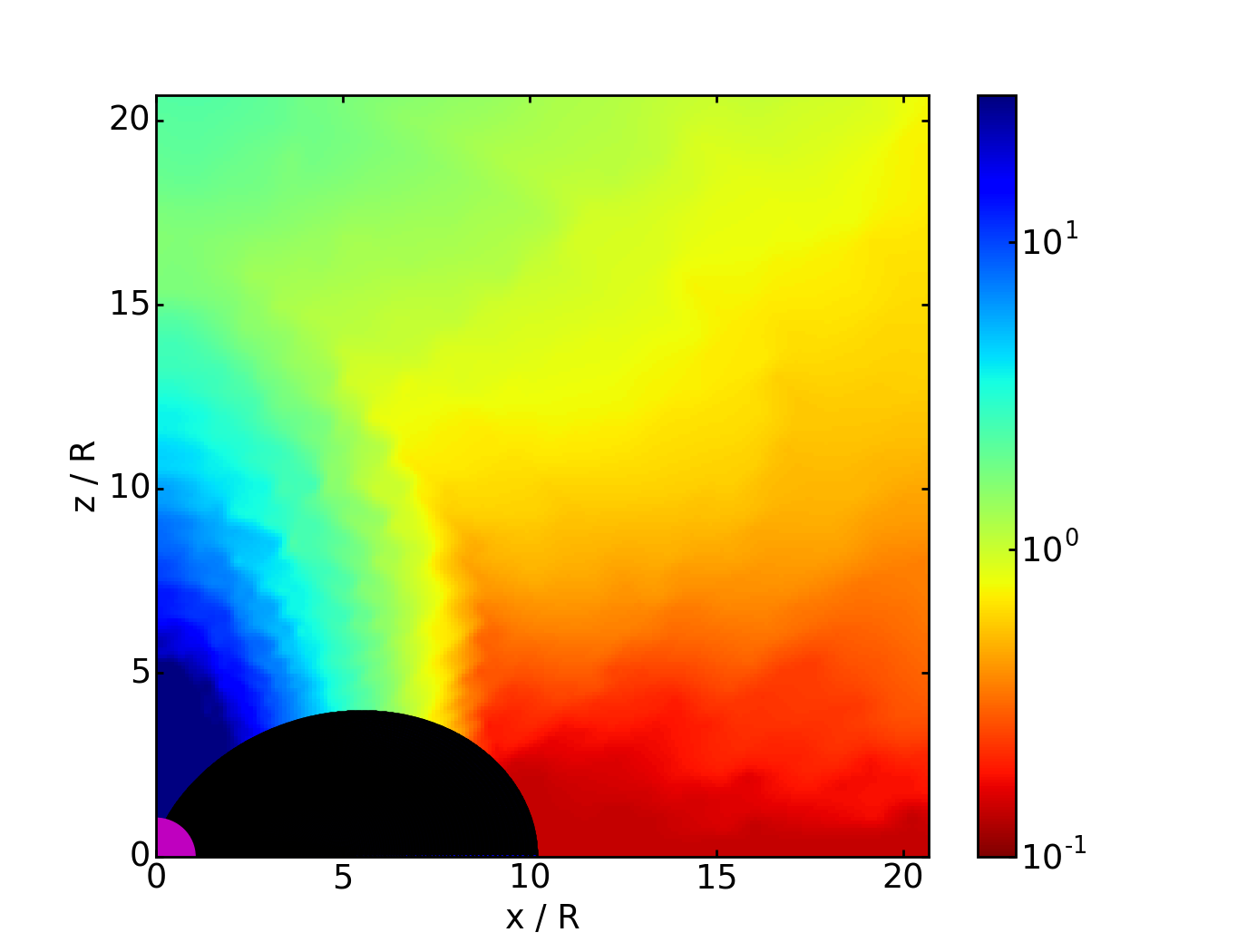}}
 {\caption{Momentum of the fastest e$^+$ of a waterbag distribution for the 
self-consistent outflow solution. The black area is a cavity 
without currents, and the magenta sphere represents the neutron 
star.}}\label{fig_mom}
\end{figure}

Assuming a waterbag distribution that separates the flow into slowly moving 
e$^-$ 
and faster moving e$^+$ with a flat momentum distribution, the flow can be 
completely described by the momentum $p_+$ of the fastest e$^+$
\citep{Beloborodov2007,Beloborodov2009,Beloborodov2013long}. For a self-similar 
magnetic field configuration with prescribed twist 
$\Delta\Phi$ \citep{Thompson2002}, one obtains a 
self-consistent solution for $p_+$ as shown in Fig.\,\ref{fig_mom}. Our result 
is consistent with that presented in \cite{Beloborodov2013long}.

\begin{figure}
 \SPIFIG{\includegraphics[width=.48\textwidth]{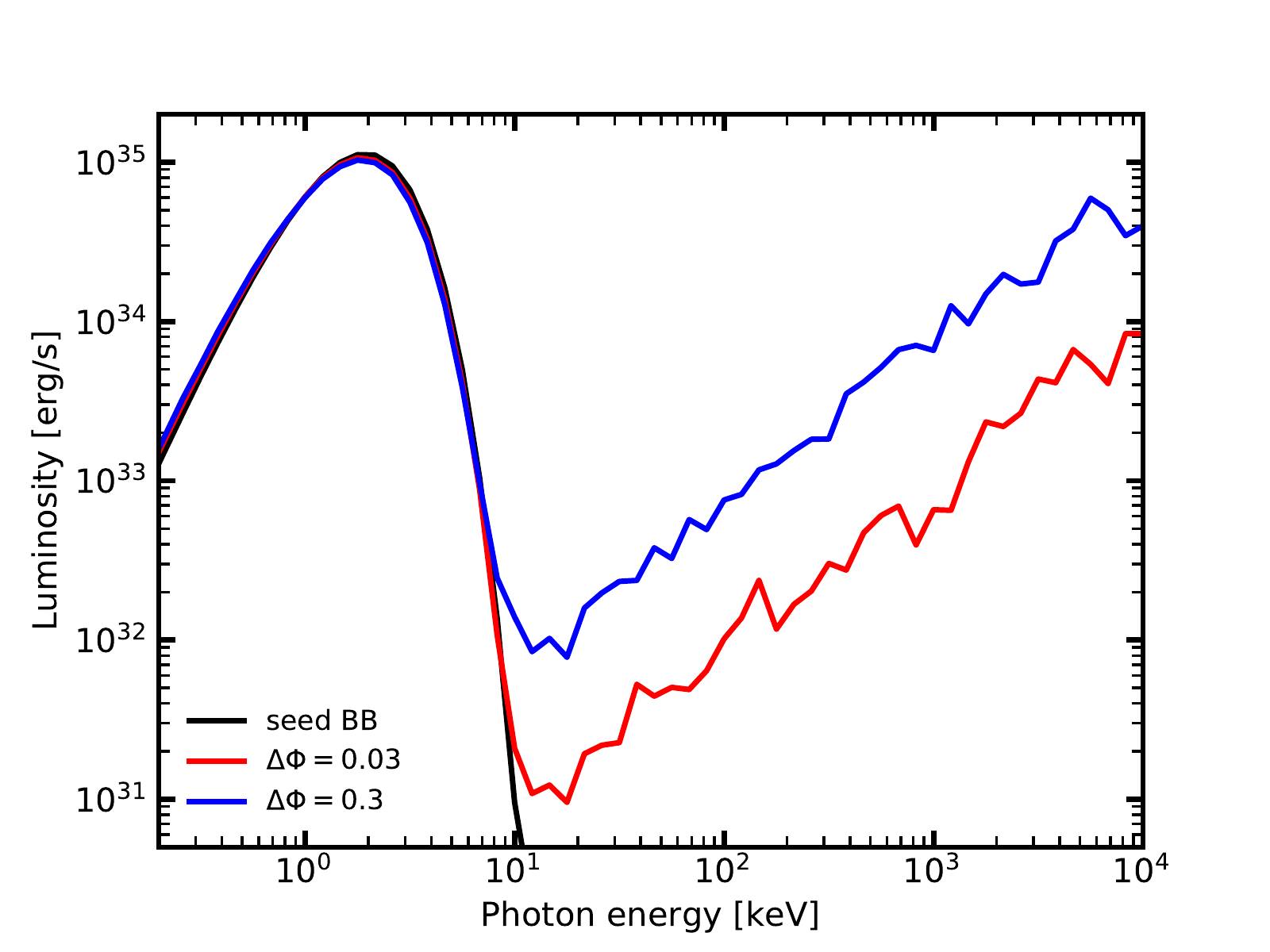}}
 {\caption{Spectra for low ($\Delta\Phi=0.03$, red line) and 
moderate ($\Delta\Phi=0.3$, blue line) twist compared to the seed black-body 
spectrum (black line). }}\label{fig_spec}
\end{figure}

The corresponding spectra for two different $\Delta\Phi$ and for $\bar 
B=10^{15}\,$G are plotted in 
Fig.\,\ref{fig_spec}. The more the magnetic field is twisted, the more photons 
get scattered to high energies. A steady twisting and untwisting causes changes 
in the light curve. For a simplified model of the momentum distribution we 
obtained estimates for the necessary surface amplitudes of less than $1\,$km 
\citep{Gabler2014}.

\section{Conclusion}\label{sec_conclusion}
We have shown that by associating the QPOs observed in giant flares of 
magnetars with torsional magneto-elastic oscillations one can constrain 
properties of high density matter. Within our model, we require realistic 
magnetic field strengths $2\times10^{14}\,\mathrm{G}\lesssim\bar 
B\lesssim5\times10^{15}\,\mathrm{G}$, and e.g.
for crust thicknesses $\Delta 
r<0.5\,$km we require shear moduli 
$\mu_{cc}\lesssim9\times10^{29}\,$erg/cm$^3$. Oscillations being
symmetric with respect to the equator can be transmitted to the magnetosphere 
creating a steadily twisting and untwisting magnetic field there. A twisted 
magnetic field requires currents flowing in the magnetosphere, which in turn 
can change the spectrum due to RCS. The changing spectrum leads to a 
modulation of the light curve of the flare. The process of the scattering is 
complicated and determined by a strong coupling between particle momenta and 
photon spectrum. 
With a simplified prescription for the particle momenta we obtain estimates 
for the surface amplitudes of the oscillations of less than $1\,$km.

\section*{Acknoweledgements}
Work supported the Spanish MINECO (grant AYA2015-66899-C2-1-P),
 the {\it Generalitat Valenciana} (PROMETEOII-2014-069), and the EU through the 
ERC Starting Grant no. 259276-CAMAP and the ERC Advanced Grant
no. 341157-COCO2CASA. Partial support comes from the COST Actions 
 NewCompStar(MP1304) and PHAROS (CA16214). Computations were performed at the 
{\it Servei  d'Inform\`atica de 
la Universitat de Val\`encia} and at the Max Planck Computing and Data Facility 
(MPCDF).

\section*{References}

\def\na{\rm{NewAstronomy}} 
\def\url#1{}
\def\doi#1{}
\bibliography{magnetar}

\end{document}